\documentclass[12pt,a4paper]{article}
\usepackage{amssymb,amsmath,graphicx,epsfig,subcaption,textcomp, xcolor,cite}
\usepackage[utf8]{inputenc}
\usepackage[english]{babel}

\begin{document}
\begin{center}{\Large {\bf Disorder effects in Ising metamagnetic phase transition} }\end{center}

\vskip 0.5cm
\begin{center}{\it  Ajanta Bhowal Acharyya$^{1,a}$ and Muktish Acharyya$^{2,b,*}$}\end{center}

\vskip 0.2cm

\begin{center}
{$^1$Department of Physics, Lady Brabourne College, Kolkata-700017, India}\\

{$^2$Department of Physics, Presidency University, Kolkata-700073, India}\\

\end{center}

\vskip 0.4cm

\noindent {\bf Abstract:} The thermodynamics of randomly quenched disordered Ising metamagnet has been studied by Monte Carlo
simulations. The disorder has been implemented either by inserting nonmagnetic impurity or by uniformly distributed quenched random
magnetic field. The staggered magnetisation ($M_s$) (calculated from the sublattice magnetisation) and the corresponding staggered susceptibility
($\chi$) are studied 
as functions of the temperature ($T$). The antiferromagnetic phase transition has been found while cooling the system from the high temperature paramagnetic phase. The transition temperature
(or pseudocritical temperature ($T_c$)) has been found to decrease as the concentration ($p$)  of nonmagnetic impurity increased.
 \textcolor{blue}{The nonmagnetic impurity dependent staggered magnetisation has been found to show the scaling behaviour $M_sp^b \sim (T-T_c)p^a$ (with $a \cong -0.95$, $b \cong 0.09$ and $T_c \cong 4.45$) obtained through the data collapse. The zero temperature staggered magnetisation ($M_s(0)$) has been found to decrease linearly. The critical temperature($T_c$) is showing 
 a linear ($T_c=mp+c$) dependence with the concentration ($p$) of nonmagnetic impurity.}
  The antiferromagnetic phase transition has been found to take place at lower temperature for the higher value of the width ($s$)
of the uniformly distributed quenched random field. \textcolor{blue}{The critical temperature ($T_c$) has been found to show
the nonlinear dependence ($T_c=a+bs+cs^2$) on the width ($s$) of the uniformly distributed random magnetic field. The extrapolation (both for $p \to 0$ and $s \to 0$) restores the Neel temperature of three dimensional pure Ising antiferromagnet.}

\vskip 4cm


\vskip 3cm

\noindent $^a${E-mail:ajanta.bhowal@gmail.com}\\
$^b${E-mail:muktish.physics@presiuniv.ac.in}\\
$^*$ Corresponding author
\newpage

\section{Introduction}
The metamagnetic systems show\cite{chaikin} the interesting tricritical behaviour with the rich phase diagrams. The experimental and theoretical
investigations are going on to explore the thermodynamic phase transitions of the metamagnetic systems. The Iron-Halide  
compounds show the metamagnetic behaviours. The specific heat anomally has been experimentally 
observed\cite{katori1} in ${\rm FeBr_2}$ metamagnet in the presence of external magnetic field. The field induced transverse ordering has 
been studied\cite{oleg1} in the same system experimentally. Very recently, the metamagnetic multiband Hall effect has been
studied experimentally \cite{kurumaji} in metamagnetic ${\rm ErGa_2}$ systems.

The experimental finding of metamagnetic behaviours in ${\rm FeBr_2}$ systems has drawn attention of the theoretical and
simulational researchers. The Monte Carlo simulation with Ising like interaction has been investigated\cite{michel1} to
resolve the above mentioned specific heat anomaly\cite{katori1}. The non-classical effects with off-diagonal interactions 
are also studied\cite{michel2}. Finally, the specific heat anomaly is resoved\cite{nowak} in Heisenberg metamagnet studied
by Monte Carlo simulation.

The metamagnets are interesting systems for other theoretical studies. Namely, the equilibrium properties of Ising metamagnetic thin films has been investigated\cite{chou}. The magnetic behaviours of Ising metamagnet with both transverse and longituninal fields are studied\cite{geng}. The behaviour of Ising metamagnet has also been studied\cite{dasgupta} in staggered field.
The tricritical behaviour of Ising metamagnet in presence of both transverse and longitundinal magnetic field has been investigated
\cite{liu}. The thermodynamic phases and the transitions are studied recently \cite{erol} in continuous symmetric spin models (XY model) by Monte Carlo simulation.

The above mentioned studies are mainly the equilibrium responses of metamagnets. The nonequilibrium responses of the metamagnetic systems have also been studied in recent past. The competition between intrinsic time scale and that of the external field has been investigated\cite{gulpinar} in Ising metamagnet. The effective field theoretical studies have drawn the attention \cite{deviren} for Ising
metamagnet driven by oscillating magnetic field. The Monte Carlo investigation for nonequilibrium responses of synthetic metamagnetic film is found\cite{mayberry} in the literature. The nonequilibrium phase transition in Ising metamagnet driven by 
oscillating ang propagating magnetic field wave have been 
studied\cite{acharyya1,acharyya2}. The results of nonequilibrium phase transition in driven Blume-Capel metamagnet has been reported
\cite{keskin1}.

The disordered metamagnetic behaviour has also been investigated\cite{godoy,santos,zukovicz}. However, the systematic study
and the thermodynamic phases and the phase diagrams of disordered metamagnets are missing in the literature. In this article,
we took the initiative of systematic Monte Carlo investigations of the thermodynamic phases and the dependence of transition
temperature on the amount of disorder. The two types of disorders have been considered here, namely, randomly distributed nonmagnetic impurity
and the uniformly distributed random field. We report the dependence of the critical temperature on the concentration of 
the nonmagnetic impurity and on the width of the uniformly distributed random magnetic field. The manuscript is organized as
follows: Section-2 is devoted to introduce the disordered Ising metamagnetic model by the corresponding Hamiltonian.
The Monte Carlo methodology is briefly mentioned in Section-3. The numerical results are given in Section-4. Section-5 
contains the concluding remarks.

\section{Model of Ising Metamagnet}


The Ising metamagnet (layered antiferromagnet) is represented by the following Hamiltonian:
\begin{equation}
\label{eq:hamiltonian}
\mathcal{H} = -J_f \sum_i \sum_{\langle ij \rangle}S_iS_j -J_a \sum_i \sum_{\langle ik \rangle}S_iS_k - \sum_{i}h_i S_i,
\end{equation} 
where the spin variables ($S_i$) are allowed to accept the value either +1 or -1. Here, $\langle ij \rangle$ indicates summation over nearest neighbors (of the spin at i-th site) only. The first term represents the ferromagnetic spin-spin interaction $(J_f > 0)$, acting among the spins in a particular layer. The interlayer interaction is antiferromagnetic ($J_a < 0)$ and is represented by second term.
The third term is the Zeeman term (interaction of spins with the site dependent random magnetic field $h_i$). The periodic boundary conditions are applied in all three directions of a solid cube of size $L$.

\section{Monte Carlo simulation scheme}

We have employed the standard and conventional Monte Carlo simulational method to study the equilibrium thermodynamic properties of this system in 
the presence of randomly quenched disorder. The randomly quenched disorder may be introduced in the system by two ways, namely
(i) inserting the nonmagnetic
impurities in random position with particular concentration (ii) by random magnetic field with particular statistical distribution (uniform here). The 
random nonmagnetic impurity may arise from the preparation of such metamagnetic crystal. The random field is the
manifestation of random crystal field, arises due to lattice distortion. We have studied the effects of random
disorder for both impurity and random fields.

The pure metamagnetic phases (in the presence of steady external magnetic field) are well studied. For zero external field
the system undergoes a continuous phase transition from high temperature paramagnetic phase to the low temperature
antiferromagnetic phase. In the present study, we have focussed on this transition in the presence of disorder.

Let us briefly discuss the Monte Carlo simulation scheme employed here. The high temperature paramagnetic phase is simulated
by considering a random distribution of up ($S_i=+1)$ and down ($S_i=-1)$ spins with equal concentrations. The total magnetisation $M_T = {1 \over L^3} \sum S_i$ vanishes for such a configuration, yielding paramagnetic phase. We have considered $L=20$ throughout the study. \textcolor{blue}{For such random configuration, the staggered magnetisation will vanish also.} However, at any 
lower temperature this should not be the equilibrium configuration. The equilibrium configuration has been achieved by
flipping the spins (chosen randomly) with Metropolis probabaility\cite{binder} of spin flip, $P(S_i \to -S_i) = {\rm Min}(1, e^{-{{\delta {\mathcal H}} \over {k_B T}}})$. The $\delta {\mathcal H}$ is the change in Hamiltonian (or energy) due to the spin flip. $k_B$ is Boltzmann constant
and $T$ is the temperature of the system. For simplicity, we have taken $k_B=1$. We have considered $J_f=1$ and
$J_a=-1$ throughout the simulation. In this way $L^3$ number of spins are updated randomly. This constitutes the unit time step and is
called Monte Carlo Step per Site (MCSS). Many such MCSS is required to achieve the equilibrium configuration at any
fixed temperature ($T$). The temperature is measured in the unit of ${J_f}/{k_B}$. The macoroscopic thermodynamic quantities are averaged over many such MCSS assuming the ergodicity which 
provides identical results of time averaging and the ensemble averaging. The system is being cooled by reducing the temperature
in a small steps ($\Delta T=0.05$ considered here). \textcolor{blue}{The final configuration of higher themperature ($T$) is used as the initial configuration of next lower temperature ($T-\Delta T$) during the cooling. In the present study, for a fixed temperature ($T$),
we have allowed the system to pass over $2\times 10^4$ MCSS and discarded $10^4$ MCSS transient steps. Hence, we have
averaged over the rest $10^4$ MCSS. The results for each temperature are further averaged over 10 different random realizations
of nonmagnetic impurities.}

\section{Simulational results} 
We have calculated the staggered magnetisation $(M_s)$ and corresponding susceptibility $\chi$. The staggered magnetisation is
defined as $M_s = (|M_o - M_e|)/2$, where $M_o={{1} \over {L^2}}\sum S_i$ (for odd plane) and $M_e={{1} \over {L^2}}\sum S_i$ (for even palnes) refer the magnetisation densities for odd and even planes respectively.
\textcolor{blue}{It may be noted here that this definition captures the right symmetry of the antiferromagnetic order parameter.
Since, the $M_o$ and $M_e$ are calculated as the magnetisation densities, the staggered magnetisation also reveals the
meaning of the staggered magnetisation densities.
As a result, the susceptibility corresponding to the satggered magnetisation is defined as , $\chi = {{L^3} \over {k_BT}}(\langle M_s^2 \rangle - \langle M_s \rangle^2)$\cite{binder}. The same definition of the susceptibility for the ferromagnet, has been employed here.} 

\begin{figure}[h!]
\centering
\includegraphics [angle=0,scale=0.85] {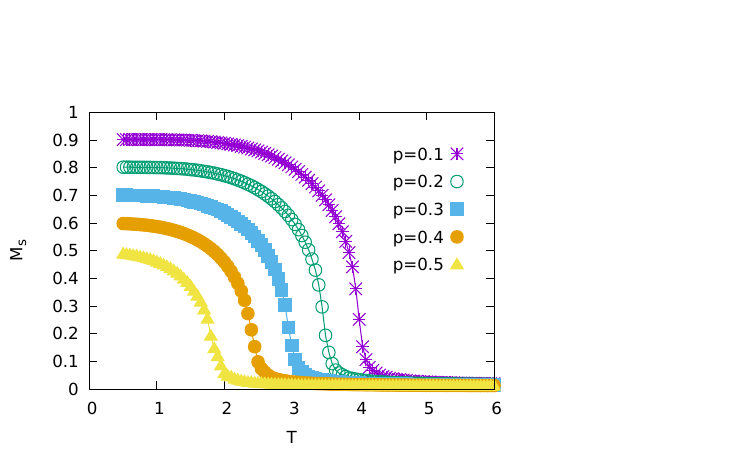}\\
a

\includegraphics [angle=0,scale=0.8] {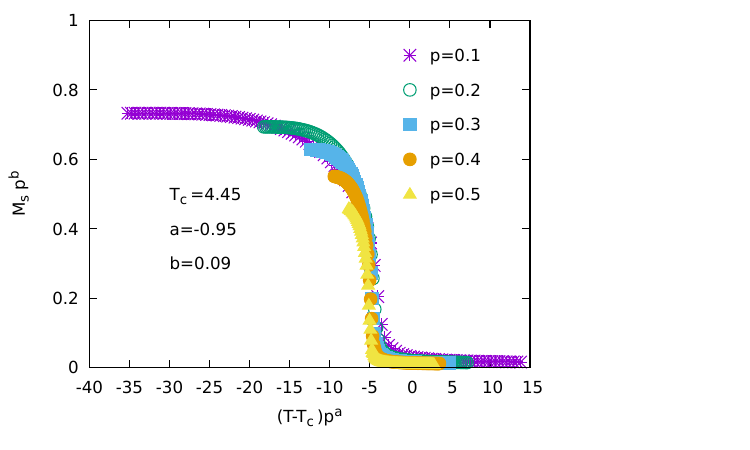}\\
b

\includegraphics [angle=0,scale=0.9] {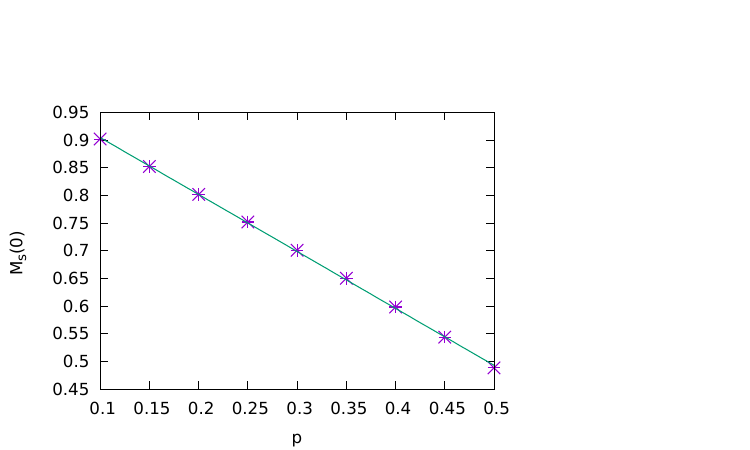}\\
c

\caption{(a) Plot of staggered magnetisation $M_s$  versus temperature ($T$) for  five
different values of impurity concentrations. (b) The scaled staggered magnetisation ($M_s p^b$) is plotted
against scaled reduced temperature ($(T-T_c)p^a$. The best estimates are $T_c=4.45$, $a=-0.95$ and $b=0.09$ for data collapse. (c) The zero temperature (extrapolated) staggered magnetisation
($M_s(0)$ is plotted against impurity concentrations ($p$). The best fitted ($M_s(0)=mp+c$) parameter values are $m=-1.029$ and $c=1.007$. Here, $h_i=0$.
  }
\label{Ms-T_p}
\end{figure}

We have studied the staggered magnetisation ($M_s$) and the susceptibility ($\chi$) as functions of the temperature ($T$) for the different values of impurity concentrations ($p$). Our system size is $L=20$. The  staggered magnetisations ($M_s$), 
are plotted (in Fig-\ref{Ms-T_p}(a)) against the temperature ($T$)  for five different values of impurity concentrations  
($p$ ). It has been observed that for any fixed value of the impurity concentration $p$, the staggered magnetisation ($M_s$)
has been found to decrease as the temperature ($T$) increased. The staggered magnetisation eventually vanishes at the transition
temperature ($T_c$). \textcolor{blue}{For any fixed value of the temperature ($T$), the staggered magnetisation ($M_s$) has been found to decrease as the concentration ($p$) of the nonmagnetic impurity increased. So the staggered magnetisation has a strong
dependence on both the temperature ($T$) as well as the concentration ($p$) of nonmagnetic impurity. The pseudocritical temperature (where the staggered magnetisation vanishes for any fixed value of impurity concentration) is also affected
by the nonmagnetic impurity. We have tried to find any possible scaling realtion among the staggered magnetisation, temperature
and the impurity concentration. By simple data collapse (shown in Fig-\ref{Ms-T_p}(b)), we have proposed the scaling behaviour $M_sp^b \sim (T-T_c)p^a$, with estimation $a \cong -0.95$, $b \cong 0.09$ and $T_c \cong 4.45$. This value of $T_c$ is comparable
with the Monte Carlo estimates ($T_c=4.511$) of the Neel temperature of simple cubic Ising antiferromagnet\cite{ferrenberg}.}

\textcolor{blue}{The zero temperature staggered magnetisation($M_s(0)$) \textcolor{blue}{(obtained by the extrapolation
$T \to 0$ from Fig-\ref{Ms-T_p}(a))} has been found to decrease as the concentration of
the nonmagnetic impurity is increased. The Fig-\ref{Ms-T_p}(c) shows the  plot of $M_s(0)$
versus  the concentration ($p$) of the nonmagnetic impurity. The best fit shows the linear relationship (
$M_s(0)=mp+c$ with $m=-1.029$ and $c=1.007$.}  

\begin{figure}[h!]
\centering
\includegraphics[angle=0,scale=1.1]{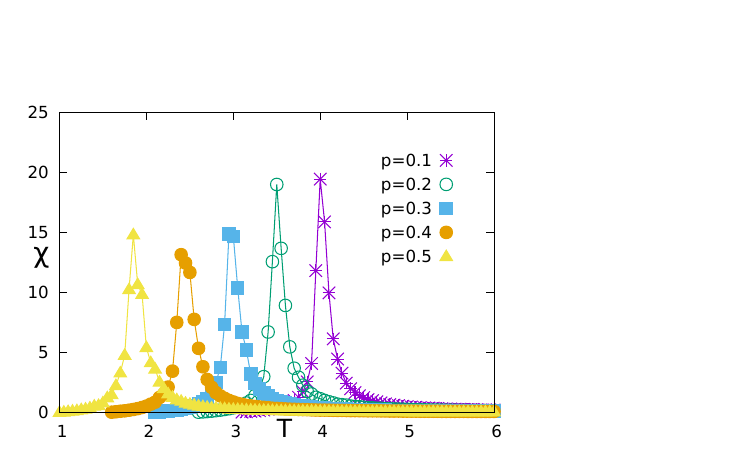}
\caption{The staggered susceptibility ($\chi$) is plotted against the temperature($T$) for  five
different values of  
the nonmagnetic impurity concentrations ($p$),  
 in the absence of any magnetic field ($h_i=0.0$). Here, the system size $L=20$.}
\label{chi-T_p}
\end{figure}

\textcolor{blue}{How does the critical temperature depend on the concentration of the nonmagnetic impurity ? For that objective, we have tried to estimate the values of the pseudocritical temperatures. The susceptibility ($\chi$) has been studied as the function of temperature ($T$) for different values of the concentration ($p$) of the nonmagnetic impurity. For fixed value of
$p$, the susceptibility shows a sharp peak (in Fig-\ref{chi-T_p}) (assuming eventual divergence in the thermodynamic limit $L \to \infty$) at the temperature which estimates the pseudocritical temperature ($T_c$).}

\begin{figure}[h!]
\centering
\includegraphics[angle=0,scale=1.1]{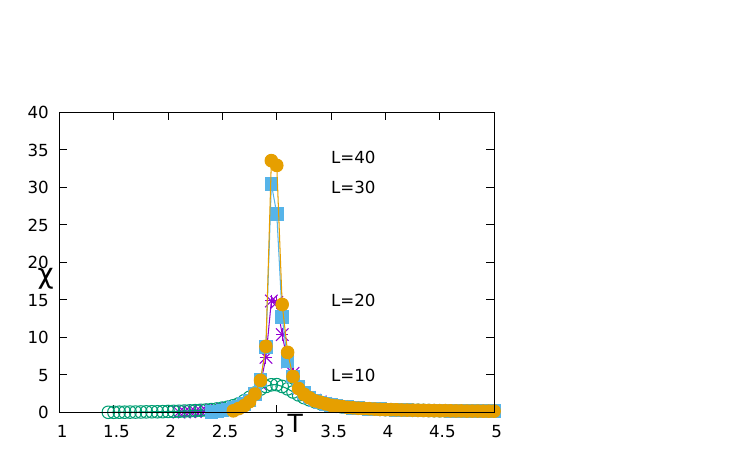}
\caption{The staggered susceptibility  $\chi$ is plotted against the temperature($T$) for  five
different values of system size ($L$) for $p=0.3$. Here, $h_i=0$.}
\label{chi-T_L}
\end{figure}

\textcolor{blue}{To confirm that the staggered susceptibility truly captures the growth of critical correlation in the thermodynamic limit, we have studied the susceptibility ($\chi$) as function of temperature ($T$) for the different values of the system size ($L=$, 10, 20, 30 and 40) and for a fixed value of
the concentration ($p=0.3$, here) of the nonmagnetic imurity. 
The plot is shown in the Fig-\ref{chi-T_L}. The height of the peak of the susceptibility grows as the system becomes larger and larger. One can reasonably expect eventual divergence (signature of the growth of critical correlation) of the susceptibility
($\chi$), a true fingerprint of conventional thermodynamic phase transition. It may be worthmentioning here that one can try to
estimate the scaling exponents for the growth of the susceptibility with the system size. However, it requires systematic
investigation with precise estimation of the height of the peak. A huge computational effort should be devoted to have such
precise estimation of the exponent. This is beyond the scope of the present study.}
 
\begin{figure}[h!]
\centering
{\includegraphics [angle=0,scale=1.2] {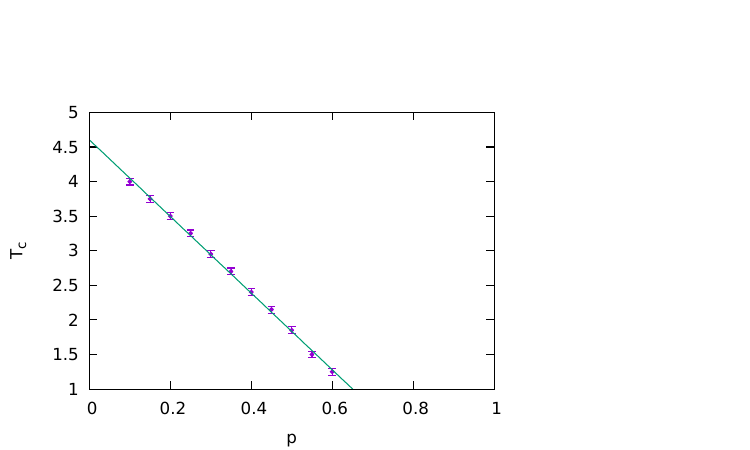}}
\caption{The critical temperature ($T_c$) is plotted against the nonmagnetic impurity concentrations ($p$), in the absence of any  magnetic field ($h_i=0.0$). 
The solid straight line represents the best fitted ($T_c=mp+c$, with $m \cong -5.536$ and $p \cong 4.601$) curve.}
\label{Tc-p} 
\end{figure}

\textcolor{blue}{We have further investigated the quantitative effects of nonmagnetic impurity on the critical temperature. The phase transition was
found to take place at lower temperature for more impure system. Here, Fig-\ref{Tc-p} depicts the results. 
Interestingly, it is  noticed that the pseudocritical temperature ($T_c$) 
decreases linearly with the increase of impurity concentration ($p$). The linear best fit suggests
 $T_c = m.p +c $, with the estimates of the fitting constants $m$ and $c$  
 as -5.536 and 4.601 respectively for minimum value of chi square. We have confirmed the goodness of fit (standard available chi square table) from the
distribution of chi square for degrees of freedom (DOF) equal to nine (we have eleven data points and two fitting parameters).
It may be noted here that for pure metamagnet ($p=0$) the, transition temperature
can be found by linear extrapolation of this curve. This value ($T_c=4.601J/k_B$) is quite close to that for the Monte Carlo estimates\cite{ferrenberg,landau} of the critical
temperature ($T_c=4.511J/k_B$) of Ising antiferromagnetic phase transition.}
Here, the nonmagnetic impurity acts as quenched disorder. The disorder compels the system to transit (antiferro-para)
at lower temperature for higher concentration. 

Now let us discuss the effects on another kind of disorder, namely, the
quenched random field. We have considered the uniformly distributed quenched random field of width $s$. The random field is uniformly
distributed between $-s/2$ and $+s/2$. We have studied the dependence of the pseudocritical temperature on the width
($s$) of the random field ($h_i$). In this case the concentration ($p$) of nomagnetic impurity is zero.

\begin{figure}[h!]
\centering
\includegraphics [angle=0,scale=1.2] {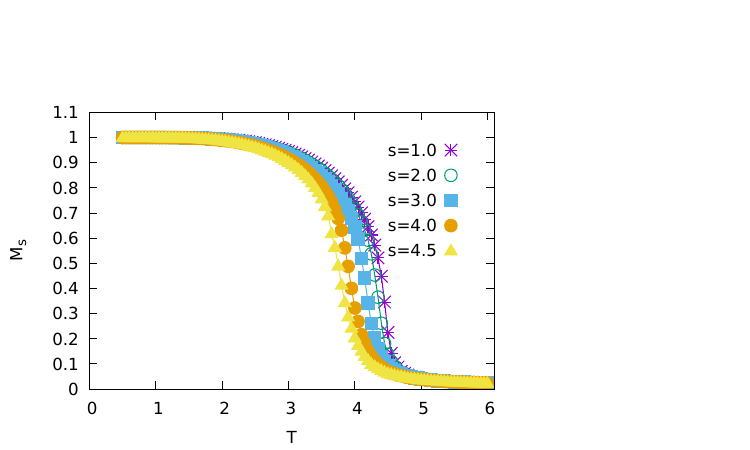}
\caption{The staggered magnetisation ($M_s$), of pure ($p=0$) Ising metamagnet, is plotted against the temperature ($T$) for  five
different values of the width ($s$) of the uniformly distributed random magnetic field. }

\label{Ms-T_s}
\end{figure}

To know the effects of random field, the staggered magnetisation ($M_s$) has been studied as function of the 
 temperature $T$  for  different values of the width ($s$) of the unformly distributed random field.
In this case, it has been observed that for fixed $s$, the staggered magnetisation ($M_s$)
decreases with temperature ($T$) and vanishes at the transition
temperature ($T_c$). The Fig-\ref{Ms-T_s} shows the  plot of $M_s$
versus temperature ($T$) for five different values of $s$.

\begin{figure}[h!]
\centering
\includegraphics[angle=0,scale=1.1]{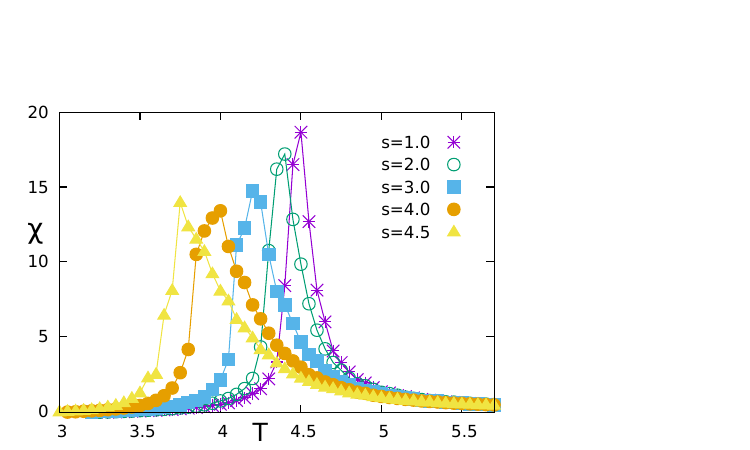}
\caption{The staggered susceptibility ($\chi$), of pure ($p=0$) Ising metamagnet, is plotted against the temperature($T$) for  five
different values of the width ($s$) of the distribution of uniformly distributed random magnetic field. Here, $L=20$.}

\label{chi-T_s}
\end{figure}

The corresponding staggered susceptibility $\chi$ has been studied also as a function of the temperature $T$ and shown in Fig-\ref{chi-T_s}.
The susceptibility gets peaked at any finite temperature which has been recognized as the pseudocritical temperature
($T_c$). Moreover, the phase transition (from paramagnetic to antiferromagnetic as the system is cooled down) has been found to take place
at lower temperature for higher values of the width ($s$) of the uniformly distributed random magnetic field.

\begin{figure}[h!]
\centering
{\includegraphics [angle=0,scale=1.3] {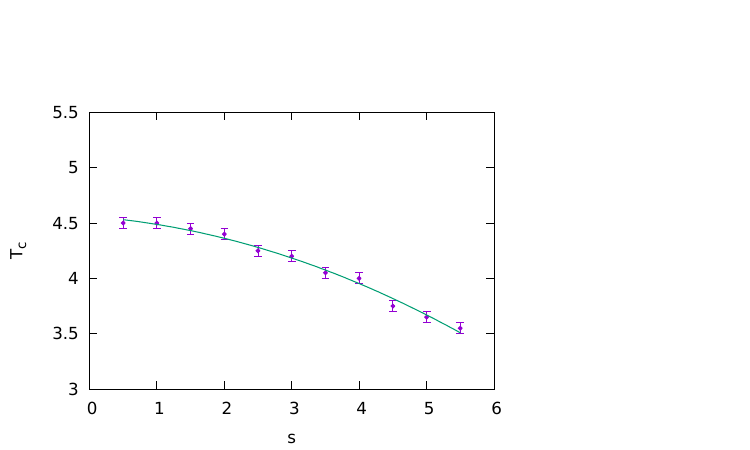}}
\caption{The critical temperature ($T_c$), of pure ($p=0$) Ising metamagnet, is shown as a function of the width ($s$) of the uniformly distributed 
random magnetic field. The solid line is the best fitted ($T_c=a+bs+cs^2$) curve with $a=4.560$, $b=-0.0465$ and $c=-0.0263$. } 
\label{Tc-s} 
\end{figure}

\textcolor{blue}{Here, unlike the case of nonmagnetic impurity, the nonlinear decrease of $T_c$ with $s$, is noticed
(shown in Fig-\ref{Tc-s}. We have fitted the data with
a nonlinear function $T_c(s)=a+bs+cs^2$ for minimum value of chi squre. The best fit estimates are $a=4.560$, $b=-0.0465$ and
$c=-0.0263$. Here also, we have confirmed the goodness of fit (standard available chi square table) from the
distribution of chi square for degrees of freedom (DOF) equal to nine (we have eleven data points and two fitting parameters).
Interestingly, in the limit of vanishingly small random field ($s \to 0$), the transition temperature
can be found by extrapolating this curve. This value ($T_c=4.560J/k_B$) is quite close to that for the Monte Carlo estimates of the critical
temperature ($T_c=4.511J/k_B$)\cite{ferrenberg,landau} of three dimensional pure Ising antiferromagnetic phase transition.}

\section{Concluding remarks}

The disordered Ising metamagnet (layered antiferromagnet) has been studied by Monte Carlo simulation. The disorder has been
implemented either by inserting nonmagnetic impurity (defects) or by random magnetic field (crystal field arises from lattice
dislocations/distortions). The staggered magnetisation has been calculated from the sublattice magnetisations. The corresponding staggered susceptibility has been calculated from the fluctuations of the staggered magnetisations. The staggered magnetisation and the susceptibility have been studied as functions of the temperature. As the system is cooled from a high temperature, an antiferromagnetic phase transition is found to take place at a pseudocritical temperature. The temperature, which maximizes the susceptibility, provides the estimate of 
the pseudocritical temperature. The pseudocritical temperature has been investigated with different concentrations of 
nonmagnetic impurity. The impure metamagnets reduces the pseudocritical or transtion temperature. The transition temperature
has been found to decrease linearly ($T_c = mp+c$) with the concentration ($p$) of nonmagnetic impurity.  \textcolor{blue}{The scaling behaviour, $Mp^b \sim (T-T_c)p^a$ (with estimation $a \cong -0.95$, $b \cong 0.09$ and $T_c \cong 4.45$) is proposed. The zero temperature staggered magnetisation was found to decrease linearly
($M_s(0)=mp+c$)
 with the concentration ($p$) of nonmagnetic impurty. The extrapolation in the direction of vanishing concentration
 ($p \to 0$) of nonmagnetic impurity leads to the Monte Carlo estimate of Neel  temperature of Ising antiferromagnet.}

The lattice distortion or dislocation produces disorder in the system. The lattice distortion or dislocation can be 
incorporated by random magnetic field on each lattice site. This random magnetic field has been considered uniformly distributed random values of finite width.
 \textcolor{blue}{The pseudocritical temperature has been studied as function of width of this uniformly distributed random field. 
Unlike the case of nonmagnetic impurity, the nonlinear decrease of critical temperature ($T_c(s)=a+bs+cs^2$), with the width ($s$) uniformly distributed  random field, is observed here. Here also, the
extrapolation towards the direction of vanishing disorder ($s \to 0$) leads to the Monte Carlo estimate of the Neel temperature
of three dimensional Ising antiferromagnet.}

It may be noted here that the peak height of the susceptibility (for a fixed value of nonmagnetic impurity concentration)
has been found to increase with larger system sizes. This is an indication of growth of critical fluctuation and eventual
divergence of susceptibility in the thermodynamic limit. The extensive finite size analysis is required to have any precise estimate of the scaling exponent. This is beyond the scope of the present study.

\textcolor{blue}{The experiments on ${\rm FeBr_2}$ metamagnet with impurity may confirm all these simulational results, particularly the scaling behaviour of staggered magnetisation.}

\vskip 0.6cm

\noindent {\bf Data availability statement:} Data may be available on reasonable request to Ajanta Bhowal Acharyya.

\vskip 0.2cm

\noindent {\bf Code availability statement:} Code may be available on reasonable request to Ajanta Bhowal Acharyya.

\vskip 0.2cm

\noindent {\bf Conflict of interest statement:} We declare that this manuscript is free from any conflict of interest.
\vskip 0.2cm

\noindent {\bf Funding statement:} No funding was received, particularly to support this work.

\vskip 0.2cm

\noindent {\bf Authors’ contributions:} Ajanta Bhowal Acharyya developed the code, prepared the figures, and wrote the manuscript.
Muktish Acharyya conceptualised the problem, analysed the results and wrote the manuscript.

\newpage 

\newpage

\newpage

\newpage
 
\newpage

\newpage

\newpage

\end{document}